# Superconductivity and Crystal Structures of $(Ba_{1-x}K_x)Fe_2As_2$ ($x = 0 - 1$)

*Marianne Rotter, Michael Pangerl, Marcus Tegel and Dirk Johrendt\**

The discovery of iron arsenide superconductors has opened new avenues for superconductivity research. After first reports on $LaFeAs(O_{1-x}F_x)$ with critical temperatures ($T_c$) of 26 - 43 K,[1, 2] even higher transition temperatures up to 55 K in $SmFeAs(O_{1-x}F_x)$ followed quickly.[3] It is meanwhile accepted, that iron arsenides represent a second class of high-$T_c$ superconductors[4] since the discovery of the cuprates in 1986.[5]

Similar to the cuprates, superconductivity in iron arsenides emerges from two-dimensional, magnetically ordered layers in the parent compound. LaFeAsO crystallizes in the ZrCuSiAs-type structure,[6] composed by alternating $(LaO)^+$ and $(FeAs)^-$ layers as shown on the left hand side of Fig. 1. Superconductivity is induced by partial oxidation (hole doping)[7] or reduction (electron doping) of the $(FeAs)^{\delta-}$ layers. Electron doping has been highly successful by substitution of oxide for fluoride or by oxide vacancies, whereas only one case of hole doped LaFeAsO has been reported so far.[8]

Recently, we proposed the ternary iron arsenide $BaFe_2As_2$[9] with the well known $ThCr_2Si_2$-type structure as a potential new parent compound.[10] Our idea based on the very similar structural and electronic conditions of this long known ternary arsenide in comparison to LaFeAsO. $BaFe_2As_2$ and LaFeAsO contain identical FeAs layers, which also have the same charge according to $Ba^{2+}[(FeAs)^-]_2$. Fig. 1 shows both structures in comparison. Accordingly, we were able to induce superconductivity at 38 K in $(Ba_{0.6}K_{0.4})Fe_2As_2$ by hole doping,[11] thus we have discovered a new family of superconducting iron arsenides. Our discovery was followed by reports on isotypic compounds with strontium ($T_c \approx 37$ K),[12, 13] calcium ($T_c \approx 20$ K),[14] and europium ($T_c = 32$ K)[15] within weeks. Meanwhile, a large part of the research on superconducting iron arsenides is focused on the ternary compounds rather than arsenide-oxides, because single-phase samples and also large single crystals are much easier to obtain.

Even though several findings suggest unconventional (non-BCS) superconductivity in iron arsenides,[16-18] the pairing mechanism is far from being clear and disputed in the physical community.[19] A generally accepted key aspect for both LaFeAsO and $BaFe_2As_2$ families is a magnetic and structural phase transition in the undoped phases at temperatures between 140 and 203 K.[10, 20-22] The tetragonal symmetry (space group *I*4/*mmm*) turns into orthorhombic (space group *Fmmm*) and antiferromagnetic spin ordering emerges immediately below these temperatures. Recently, we have proved the antiferromagnetic spin structure of $BaFe_2As_2$ by single crystal neutron diffraction.[23]

[\*]    Prof. Dr. Dirk Johrendt, M. Sc. Marianne Rotter,
       B.Sc. Michael Pangerl, Dipl.-Chem. Marcus Tegel
       Department Chemie und Biochemie
       Ludwig-Maximilians-Universität München
       Butenandtstr. 5-13 (Haus D), 81377 München, Germany
       Fax: (+)89 2180 77431
       E-mail: Johrendt@lmu.de
       Web: www.cup.uni-muenchen.de/ac/johrendt/index.html

[\*\*]   We thank Dr. Joachim Deisenhofer for susceptibility
       measurements. This work was financially supported by the
       DFG.

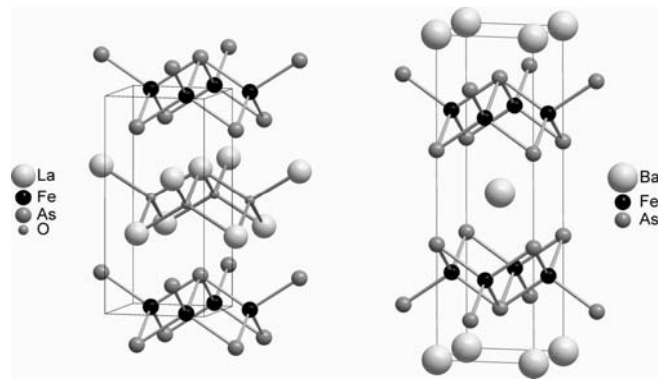

**Figure 1.** Crystal structures of LaFeAsO (left) and $BaFe_2As_2$ (right)

It has been believed that the magnetic and structural phase transitions of LaFeAsO and $BaFe_2As_2$ must be suppressed by doping, so that superconductivity can emerge. But recently it was reported, that superconductivity up to 29 K occurs even in undoped $AFe_2As_2$ ($A$ = Ca, Sr, Ba) under pressure.[24, 25] Thus all the signs are that destabilization of the antiferromagnetic state by doping or pressure is a main issue for superconductivity in iron arsenides. However, the crucial question, whether both states may coexist, is still open.

The doping dependency of the structure and superconductivity has been intensively studied on LaFeAsO-type compounds. In electron-doped $REFeAsO_{1-x}$ ($RE$ = La - Sm),[26] $T_c$ increases with higher doping levels and with decreasing lattice parameters. On the other hand, the hole doped system $(La_{1-x}Sr_x)FeAsO$[27] shows also increasing $T_c$ with higher doping levels, but with increasing lattice parameters. This indicates that the doping level is the determining parameter for $T_c$ in LaFeAsO compounds. However, these results are limited by the facts, that the exact doping levels are unknown in most cases and confined to $x \approx 0.2$ anyway. Furthermore, the changes in the lattice parameters are very small and their significance often doubtful.

In contrast to this, the $BaFe_2As_2$ system opens a golden opportunity for doping studies. This is because also $KFe_2As_2$ had been known to exist [28] and consequently K-doping of $BaFe_2As_2$ should be easy due to similar ionic radii of $Ba^{2+}$ (1.42 Å) and $K^+$ (1.51 Å).[29] So far, we have only reported on the occurrence of superconductivity in $(Ba_{0.6}K_{0.4})Fe_2As_2$. But the dependencies of superconductivity and crystal structures on the potassium content in the whole solid solution $(Ba_{1-x}K_x)Fe_2As_2$ are still unknown. In this communication, we report on the synthesis, crystal structures and superconducting transition temperatures of the complete series $(Ba_{1-x}K_x)Fe_2As_2$ ($x = 0 - 1$).

The ternary compounds $BaFe_2As_2$ and $KFe_2As_2$ both belong to the more two-dimensional branch of the $ThCr_2Si_2$-type structure without As−As bonds between the layers. The unit cells almost have the same volumes despite the slightly bigger radius of $K^+$. On the other hand, the $c/a$ ratios differ considerably because the $c$ lattice parameter of $KFe_2As_2$ is almost 1 Å longer than the $c$ of $BaFe_2As_2$. In other words, the unit cell of $KFe_2As_2$ is stretched along $c$.

The crystal structures of the compounds $(Ba_{1-x}K_x)Fe_2As_2$ were determined by Rietveld refinements of X-ray powder patterns as shown exemplary in Figure 2. Figure 3 shows the changes of the structure with doping. The lattice parameters $a$ and $c$ vary linearly with the potassium content over the whole range. We find a constant unit cell volume within the experimental error, because the strong elongation of $c$ is almost compensated by the decrease of $a$.[30] Also the Fe−As and Ba(K)−As bond lengths remain unchanged. Both parameters vary by less than 0.4% and are therefore not shown.

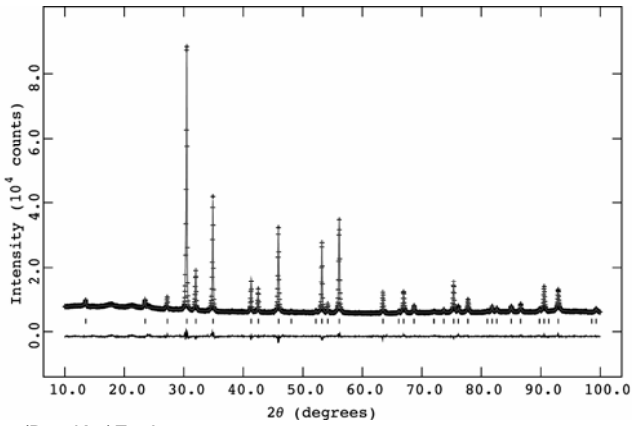

**Figure 2.** Measured (+) and calculated (—) X-ray powder pattern of $(Ba_{0.9}K_{0.1})Fe_2As_2$.

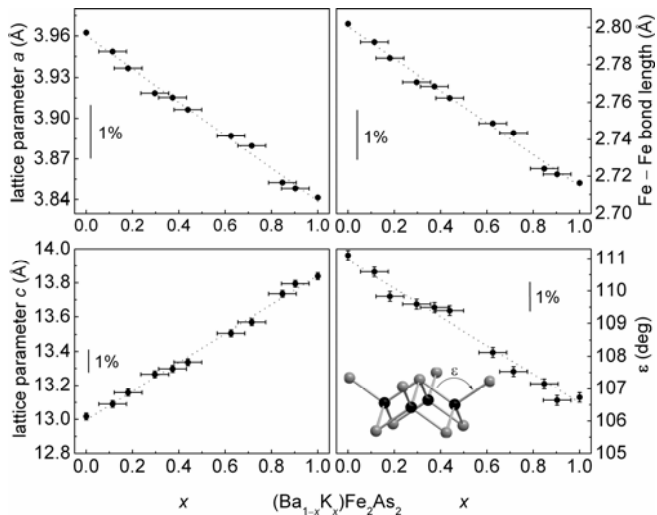

**Figure 3.** Variation of structural parameters in $(Ba_{1-x}K_x)Fe_2As_2$.

Apart from the lattice parameters, only the Fe−Fe bond length and the As-Fe-As bond angle ε changes significantly (by 5-7%) on doping. Both decrease linearly with increasing potassium content, which means that the $FeAs_4$ - tetrahedra get more elongated along $c$ and the iron atoms within the layers move together. Interestingly, ε becomes the ideal tetrahedral angle of 109.5° at $x \approx 0.4$. The insert in Figure 3 depicts the ε angle in the FeAs layer. Thus, the main implication of doping on the crystal structure of $(Ba_{1-x}K_x)Fe_2As_2$ is a decreasing As-Fe-As bond angle and a shortening of the distances between the iron atoms at the same time.

Chemical bonding in $ThCr_2Si_2$-type compounds has been intensively studied.[31] We have shown that the properties of these compounds depend on a subtle balance between different bonding interactions, especially on the interplay between metal-ligand (Fe−As) and metal-metal (Fe−Fe) bonding within the layers.[32] In the case of $BaFe_2As_2$, it is accepted that the Fe-$3d_{x2-y2}$-orbitals close the Fermi level play a key role for magnetism and superconductivity. The angle ε determines the overlap between Fe-$3d_{x2-y2}$ and As-$3sp$ orbitals, thus our results suggest a strong coupling of structural and electronic degrees of freedom by doping.

It is disputed if the structural phase transition in the iron arsenides has to be completely suppressed before superconductivity emerges. Recent results suggest that the structural distortion of LaFeAsO disappears by doping exactly at the border to the superconducting state.[33] In the case of $BaFe_2As_2$, we have already shown that the tetragonal to orthorhombic phase transition is suppressed in $(Ba_{0.6}K_{0.4})Fe_2As_2$.[11] In order to delimit the composition range of the transition, we have measured X-ray powder diffraction patterns of $(Ba_{1-x}K_x)Fe_2As_2$ ($x = 0 – 0.3$) between 300 and 10 K. Figure 4 shows the temperature dependencies of the (110)-reflections. The reduction of the lattice symmetry is visible by peak splitting or broadening up to $x = 0.2$, but absent at $x = 0.3$. The transition temperatures ($T_{tr}$) decrease strongly with higher potassium content from 140 K to ≈ 90 K at $x = 0.2$, where the transition proceeds over a wide temperature range. From this we conclude that the orthorhombic phase (space group $Fmmm$) exists at low temperatures up to $x = 0.2$ and becomes tetragonal between $x = 0.2$ and 0.3.

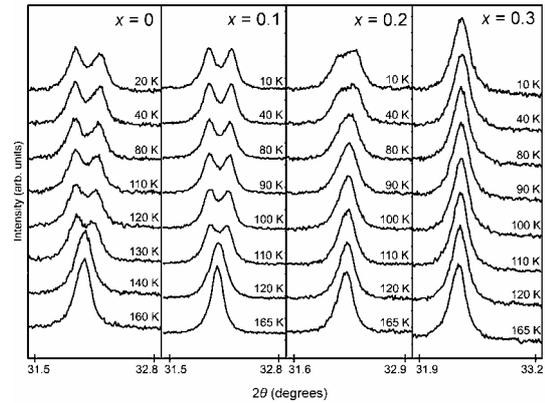

**Figure 4.** Temperature dependencies of the (110)-reflections in $(Ba_{1-x}K_x)Fe_2As_2$ with x = 0 - 0.3.

We have also investigated the doping effect on the superconducting transition temperatures. Therefore we have measured the electrical resistances of $(Ba_{1-x}K_x)Fe_2As_2$ samples ($x = 0 - 1$) between 2 and 300 K by a four probe method. The relative changes of the resistance with temperature ($R/R_{300K}$) of all samples are shown in Figure 5. Superconductivity was detected in all cases except for the undoped parent compound $BaFe_2As_2$, but the transition temperatures vary strongly. $BaFe_2As_2$ is a poor metal with a specific resistivity around 1 mΩcm at room temperature and exhibits the structural and magnetic phase transition at 140 K,[10] which is clearly visible in the resistance plot.

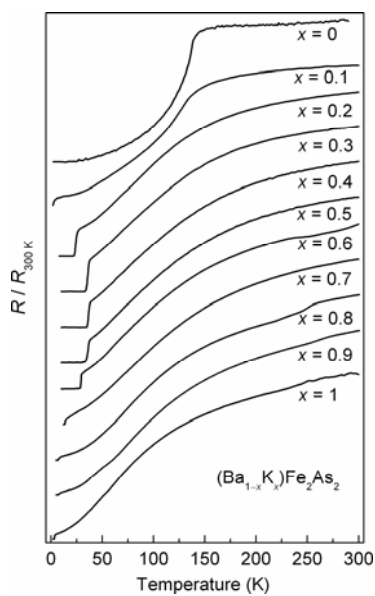

**Figure 5.** Relative electrical resistances of $(Ba_{1-x}K_x)Fe_2As_2$ samples.

At the smallest doping level $x \approx 0.1$, the resistance anomaly is less pronounced, but not completely suppressed. We find an abrupt drop in the resistance at $\approx 3$ K, which is the onset of superconductivity. However, zero resistance could not be achieved at 1.8 K, but superconductivity was proved by magnetic measurements. The anomaly in the resistance appears to be suppressed when the doping level is at 0.2. At this point, we find the behavior of a normal metal and superconductivity at $T_c \approx 25$ K, which increases strongly to 36 K and 38 K at $x = 0.3$ and $x = 0.4$, respectively. Doping levels of $x > 0.5$ lead to a continuous decrease of $T_c$ down to 3.8 K for $KFe_2As_2$.

The phase diagram in Figure 6 shows the superconducting critical temperatures ($T_c$), as well as the phase transition temperatures ($T_{tr}$) of $(Ba_{1-x}K_x)Fe_2As_2$. We find superconductivity in the range between $x = 0.1 - 1.0$ with $T_c > 30$ K between $x = 0.3 - 0.6$ and a maximum of 38 K close to $x = 0.4$. The orthorhombically distorted crystal structure exists up to $x = 0.2$, that is the point where $T_c$ is already 25 K. Thus, superconductivity apparently coexists with the orthorhombic structure and potentially with the antiferromagnetic state below $T_{tr}$ known for $BaFe_2As_2$.

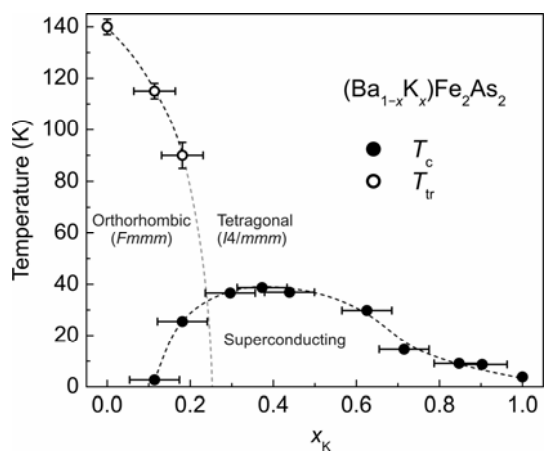

**Figure 6.** Phase diagram of $(Ba_{1-x}K_x)Fe_2As_2$ with the superconducting ($T_c$) and phase transition ($T_{tr}$) temperatures. $T_c$ is defined as the temperature where the resistance is dropped to 90% of the extrapolated value. The dashed lines are guides for the eye.

In summary, we have shown experimental doping dependencies of the crystal structure and superconductivity in the solid solution $(Ba_{1-x}K_x)Fe_2As_2$. As the main effect of doping on the crystal structure at room temperature, we find linear decreasing As-Fe-As bond angles ($\varepsilon$) and Fe–Fe distances, equivalent to an elongation of the $FeAs_4$ tetrahedra along [001]. This contradicts results of recently reported DFT calculations, where the opposing effect was proposed.[34] The structural changes are intimately coupled to the electronic states at the Fermi level, because the most relevant Fe-$3d_{x^2-y^2}$ orbitals are strongly affected by the bond angle $\varepsilon$. We have observed the structural phase transition ($I4/mmm \rightarrow Fmmm$) of $BaFe_2As_2$ with decreasing transition temperatures up to a doping level of $x = 0.2$ and the complete suppression at $x = 0.3$. Superconductivity occurs over the whole doping range in $(Ba_{1-x}K_x)Fe_2As_2$ with a maximum $T_c$ of 38 K at $x \approx 0.4$. Solely the parent compound $BaFe_2As_2$ is non-superconducting above 1.8 K. The superconducting transitions in the orthorhombic compounds $(Ba_{0.9}K_{0.1})Fe_2As_2$ ($T_c \approx 3$ K) and $(Ba_{0.8}K_{0.2})Fe_2As_2$ ($T_c \approx 25$ K) give strong evidence for the coexistence of superconductivity with the structurally distorted and potentially magnetically ordered state in the $BaFe_2As_2$ family of iron arsenide superconductors.

## Experimental Section

Polycrystalline samples of $(Ba_{1-x}K_x)Fe_2As_2$ with $x = 0 - 1$ in steps of 0.1 were synthesized by heating stoichiometric mixtures of the elements (purity > 99.9%) at 823 - 1223 K in alumina crucibles enclosed in silica ampoules under argon atmosphere. In order to reduce the loss of potassium by evaporation, the gas volume was reduced by alumina inlays in the crucibles. The products were black metallic powders and stable in air for weeks. Sampling EDX measurements showed homogenous distributions of barium and potassium (±5%) and confirmed the compositions obtained from the Rietveld fits. X-ray powder diffraction patterns were recorded between 10 and 300 K using a Huber G670 imaging plate detector ($Cu_{K\alpha1}$-radiation, Ge(111)-monochromator), equipped with a closed-cycle He-cryostat. Patterns at room temperature were indexed with tetragonal body-centered unit cells according to the $ThCr_2Si_2$ type ($I4/mmm$) or with orthorhombic face-centered unit cells ($Fmmm$, $a_{ortho} \approx \sqrt{2}\, a_{tetra} - \delta$, $b_{ortho} \approx \sqrt{2}\, b_{tetra} + \delta$, $c_{ortho} \approx c_{tetra}$) at low temperatures. Small amounts of FeAs were detected as impurity phase in some samples. The crystal structures were refined by the Rietveld method using the GSAS[35] software package using Thompson-Cox-Hastings functions with asymmetry corrections as reflection profiles.[36] Electrical resistances were measured by the four probe method on cold pressed and sintered pellets (1123 K) using a He-closed-cycle refrigerator. Gold wires were fixed to the sample by silver conduction paint.

[1]  Y. Kamihara, T. Watanabe, M. Hirano, H. Hosono, *J. Am. Chem. Soc.* **2008**, *130*, 3296.
[2]  H. Takahashi, K. Igawa, K. Arii, Y. Kamihara, M. Hirano, H. Hosono, *Nature* **2008**, *453*, 376.
[3]  Z.-A. Ren, W. Lu, J. Yang, W. Yi, X.-L. Shen, Z.-C. Li, G.-C. Che, X.-L. Dong, L.-L. Sun, F. Zhou, Z.-X. Zhao, *Chin. Phys. Lett.* **2008**, *25*, 2215.
[4]  D. Johrendt, R. Pöttgen, *Angew. Chem. Int. Ed.* **2008**, *47*, 4782.

[5] J. G. Bednorz, K. A. Müller, *Z. Phys. B: Condens. Matter* **1986**, *64*, 189.
[6] V. Johnson, W. Jeitschko, *J. Solid State Chem.* **1974**, *11*, 161.
[7] The term "doping" is commonly used by the physical community to express changes in the electron count in superconductors like $YBa_2Cu_3O_{7-x}$ and also in other materials. The "doping levels" are arbitrary and often much larger than in doped semiconductors, where the term doping has its seeds. In order to avoid confusion, we comply with this diction.
[8] H.-H. Wen, G. Mu, L. Fang, H. Yang, X. Zhu, *Europhys. Lett.* **2008**, *82*, 17009.
[9] M. Pfisterer, G. Nagorsen, *Z. Naturforsch. B: Chem. Sci.* **1980**, *35*, 703.
[10] M. Rotter, M. Tegel, I. Schellenberg, W. Hermes, R. Pöttgen, D. Johrendt, *Phys. Rev. B* **2008**, *78*, 020503(R).
[11] M. Rotter, M. Tegel, D. Johrendt, *Phys. Rev. Lett.*, in press *arxiv:0805.4630* **2008**
[12] G. F. Chen, Z. Li, G. Li, W. Z. Hu, J. Dong, X. D. Zhang, P. Zheng, N. L. Wang, J. L. Luo, *arxiv:0806.1209* **2008**.
[13] K. Sasmal, B. Lv, B. Lorenz, A. Guloy, F. Chen, Y. Xue, C. W. Chu, *arxiv:0806.1301* **2008**.
[14] G. Wu, H. Chen, T. Wu, Y. L. Xie, Y. J. Yan, R. H. Liu, X. F. Wang, J. J. Ying, X. H. Chen, *arxiv:0806.4279* **2008**.
[15] H. S. Jeevan, Z. Hossain, C. Geibel, P. Gegenwart, *arxiv:0807.2530* **2008**.
[16] H. Luetkens, H.-H. Klauss, R. Khasanov, A. Amato, R. Klingeler, I. Hellmann, N. Leps, A. Kondrat, C.Hess, A. Köhler, G. Behr, J. Werner, B. Büchner, *arxiv:0804.3115* **2008**.
[17] Y. Nakai, K. Ishida, Y. Kamihara, M. Hirano, H. Hosono, *J. Phys. Soc. Jpn.* **2008**, *77*, 073701.
[18] I. I. Mazin, D. J. Singh, M. D. Johannes, M. H. Du, *arxiv:0803.2740* **2008**.
[19] I. I. Mazin, M. D. Johannes, *arxiv:0807.3737* **2008**.
[20] C. d. l. Cruz, Q. Huang, J. W. Lynn, J. Li, W. Ratcliff II, J. L. Zarestky, H. A. Mook, G. F. Chen, J. L. Luo, N. L. Wang, P. Dai, *Nature* **2008**, *453*, 899.
[21] T. Nomura, S. W. Kim, Y. Kamihara, M. Hirano, P. V. Sushko, K. Kato, M. Takata, A. L. Shluger, H. Hosono, *arXiv:0804.3569* **2008**.
[22] M. Tegel, M. Rotter, V. Weiss, F. M. Schappacher, R. Pöttgen, D. Johrendt, *arXiv:0806.4782* **2008**.
[23] Y. Su, P. Link, A. Schneidewind, T. Wolf, Y. Xiao, R. Mittal, M. Rotter, D. Johrendt, T. Brueckel, M. Loewenhaupt, *arxiv:0807.1743* **2008**.
[24] P. L. Alireza, J. Gillett, Y. T. C. Ko, S. E. Sebastian, G. G. Lonzarich, *arXiv:0807.1896* **2008**.
[25] A. Kreyssig, M. A. Green, Y. Lee, G. D. Samolyuk, P. Zajdel, J. W. Lynn, S. L. Bud'ko, M. S. Torikachvili, S. N. N. Ni, J. Leao, S. J. Poulton, D. N. Argyriou, B. N. Harmon, P. C. Canfield, R. J. McQueeney, A. I. Goldman, *arXiv:0807.3032* **2008**.
[26] Z.-A. Ren, G.-C. Che, X.-L. Dong, J. Yang, W. Lu, W. Yi, X.-L. Shen, Z.-C. Li, L.-L. Sun, F. Zhou, Z.-X. Zhao, *Europhys. Lett.* **2008**, *83*, 17002.
[27] G. Mu, L. Fang, H. Yang, X. Zhu, P. Cheng, H.-H. Wen, *arxiv:0806.2104* **2008**.
[28] S. Rozsa, H. U. Schuster, *Z. Naturforsch. B: Chem. Sci.* **1981**, *36*, 1668.
[29] R. D. Shannon, C. T. Prewitt, *Acta Crystallogr., Sect. B: Struct. Sci.* **1969**, *25*, 925.
[30] Strictly speaking, the volume passes through a maximum at $x = 0.5$, but the change is only $\approx 0.1\%$.
[31] C. Zheng, R. Hoffmann, *J. Solid State Chem.* **1988**, *72*, 58-71; *J. Phys. Chem.* **1985**, *89*, 4175-4181.
[32] D. Johrendt, C. Felser, O. Jepsen, O. K. Andersen, A. Mewis, J. Rouxel, *J. Solid State Chem.* **1997**, *130*, 254.
[33] H. Luetkens, H.-H. Klauss, M. Kraken, F. J. Litterst, T. Dellmann, R. Klingeler, C. Hess, R. Khasanov, A. Amato, C. Baines, J. Hamann-Borrero, N. Leps, A. Kondrat, G. Behr, J. Werner, B. Büchner, *arxiv*:0806.3533.
[34] D. J. Singh, *arxiv:0807.2643* **2008**.
[35] A. C. Larson, R. B. Von Dreele, in Los Alamos National Laboratory Report LAUR 86-748, **2004**.
[36] L. W. Finger, D. E. Cox, A. P. Jephcoat, *J. Appl. Crystallogr.* **1992**, *27*, 79.

**Entry for the Table of Contents**

*Superconductivity*

Marianne Rotter, Michael Pangerl, Marcus Tegel and Dirk Johrendt*
__________ **Page – Page**

Superconductivity and Crystal Structures of $(Ba_{1-x}K_x)Fe_2As_2$ ($x$ = 0 - 1)

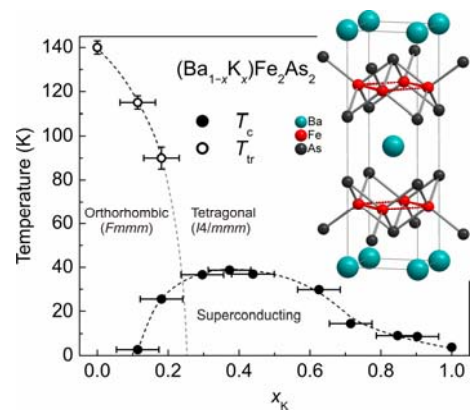

The iron arsenides $(Ba_{1-x}K_x)Fe_2As_2$ with the $ThCr_2Si_2$-type structure exhibit superconductivity between 3 K and 38 K depending on the doping level. Superconductivity appears before the structural distortion of the parent compound $BaFe_2As_2$ is completely suppressed by doping. The bond angles in the iron arsenide layers decrease by doping, suggestive of a coupling of structural and electronic degrees of freedom.